\documentclass[conference]{IEEEtran}
\IEEEoverridecommandlockouts

\usepackage[utf8]{inputenc}
\usepackage[T1]{fontenc}
\usepackage{textcomp}
\usepackage{cite}
\usepackage{amsmath,amssymb,amsfonts}
\usepackage{algorithmic}
\usepackage{graphicx,subcaption}
\usepackage{xcolor}
\usepackage{soul}
\usepackage{float}
\usepackage{stfloats}
\usepackage{tikz, pgfplots}
\usepackage[utf8]{inputenc}
\usetikzlibrary{shapes, arrows, positioning, fit, calc, backgrounds}
\usepackage[shortcuts,acronym]{glossaries}
\usepackage{booktabs}
\usepackage{balance}
\usepackage[bookmarks=false]{hyperref}

\def\BibTeX{{\rm B\kern-.05em{\sc i\kern-.025em b}\kern-.08em
    T\kern-.1667em\lower.7ex\hbox{E}\kern-.125emX}}
\begin{document}

\title{
Ray Tracing-Enabled Digital Twin for RIS Phase Optimization: Implementation and Experimental Validation
}

\author{\IEEEauthorblockN{Ömer Lütfü Karakelle\IEEEauthorrefmark{1}\IEEEauthorrefmark{4}, Sefa Kayrakl\i k\IEEEauthorrefmark{1}\IEEEauthorrefmark{3}, \.{I}brahim Hökelek\IEEEauthorrefmark{1}\IEEEauthorrefmark{4}, Ali Görçin\IEEEauthorrefmark{1}\IEEEauthorrefmark{4}, Halim Yanikomeroglu\IEEEauthorrefmark{5}}

\IEEEauthorblockA{\IEEEauthorrefmark{1} \href{https://hisar.bilgem.tubitak.gov.tr/en/} {Communications and Signal Processing Research (H\.{I}SAR) Lab., T{\"{U}}B{\.{I}}TAK B{\.{I}}LGEM, Kocaeli, Türkiye}}
\IEEEauthorblockA{\IEEEauthorrefmark{4} Department of Electronics and Communication Engineering, Istanbul Technical University, {\.{I}}stanbul, Türkiye}
\IEEEauthorblockA{\IEEEauthorrefmark{3} Department of Electrical and Electronics Engineering, Koç University, {\.{I}}stanbul, Türkiye} 
\IEEEauthorblockA{\IEEEauthorrefmark{5} Non-Terrestrial Networks (NTN) Lab, Systems and Computer Engineering, Carleton University, Ottawa, ON, Canada} 
Emails: \{omer.karakelle, sefa.kayraklik, ibrahim.hokelek\}@tubitak.gov.tr, aligorcin@itu.edu.tr, halim@sce.carleton.ca}

\newacronym{ntn}{NTN}{non-terrestrial networks}
\newacronym{tn}{TN}{terrestrial networks}
\newacronym{nb-iot}{NB-IoT}{narrowband Internet of things}
\newacronym{iot}{IoT}{Internet of things}
\newacronym{leo}{LEO}{low Earth orbit}
\newacronym{ue}{UE}{user equipment}
\newacronym{poc}{PoC}{proof-of-concept}
\newacronym{cfo}{CFO}{carrier frequency offset}

\maketitle
\begin{abstract}
Determining the optimal phase configurations of reconfigurable intelligent surface (RIS) elements typically requires complex channel estimation procedures with high pilot overhead, creating a bottleneck for real-time deployment in time-varying wireless environments. In this paper, we propose a digital twin (DT)-driven framework for RIS phase shift optimization that eliminates extensive signaling overhead associated with estimating high-dimensional RIS channels. Leveraging the NVIDIA Sionna ray-tracing library, we construct a DT of the physical environment based on a three-dimensional map. The proposed system utilizes the location information of the transceivers to compute the optimal RIS phase shift configurations within the DT. These computationally generated configurations are then transferred to a physical RIS prototype. Experimental results demonstrate that the phase configurations obtained from the DT significantly enhance the received signal power in the physical environment, validating the fidelity of the ray-tracing model and the feasibility of the proposed optimization strategy. 
\end{abstract}
\section{Introduction}

As wireless communication networks evolve towards the sixth-generation (6G), the fundamental design paradigm is shifting from optimizing the transmitter and receiver endpoints to actively controlling the propagation environment itself. In this context, reconfigurable intelligent surfaces (RISs) have emerged as a promising technology to enhance wireless connectivity. Composed of a large number of low-cost, passive reflecting elements, an RIS can manipulate electromagnetic waves and enhance coverage by inducing controllable phase shifts, thereby redirecting signals to bypass blockages. This ability to reconfigure the wireless channel in real-time offers a solution to overcome severe path loss and blockage issues by constructing a virtual line-of-sight \cite{first_par}.

Despite these promising advantages, transforming RIS-aided communication from theory to practice faces a critical bottleneck: the acquisition of accurate channel state information (CSI) for RIS-assisted paths. While theoretical studies often assume perfect CSI, obtaining it in real-world scenarios is challenging since typical RIS elements are passive, lack dedicated RF chains, and cannot generate or process pilot signals directly \cite{gp1, gp3}. Standard channel estimation techniques for RIS channels often rely on an element-wise ON/OFF switching strategy, where the receiver measures the channel response while the RIS iteratively cycles through configurations \cite{csi_est1, csi_est2}. However, this approach imposes prohibitively high pilot and timing overheads that increase linearly with the number of reflecting elements \cite{gp2, gp4}, resulting in excessive time consumption. To address these limitations, several studies in the literature have developed channel estimation methods specifically to reduce the required number of measurements \cite{omp}. While these approaches reduce the pilot overhead, they still rely on over-the-air training phases, which consume valuable time resources and yield lower performance when the number of measurements is restricted or strict time constraints are imposed. Therefore, developing low-overhead optimization strategies that bypass the need for explicit RIS channel estimation is essential for the feasible deployment of the RIS.

To mitigate the excessive overhead associated with real-time CSI estimation for RIS channels, recent research has pivoted towards location- and environment-aware communication paradigms as a promising alternative \cite{gp5, gp6}. Using the physical coordinates of the transceivers and the spatial layout of the propagation environment, these approaches allow the computation of phase shifts based on geometric information, thereby bypassing the need for extensive pilot transmission. However, prior works primarily rely on numerical simulations to evaluate these environment-aware schemes, and the experimental validation of such location-based optimization frameworks on physical RIS testbeds remains limited.

On the other hand, ray tracing (RT) has emerged as a powerful enabler for realizing environment-aware digital twins (DT) in 6G networks \cite{gp7}. Unlike statistical channel models, RT deterministically simulates the propagation of electromagnetic waves by interacting with the three-dimensional (3D) geometry of the scene, thus providing highly accurate site-specific channel data. Among the RT tools, the introduction of the NVIDIA Sionna library has accelerated this paradigm by enabling GPU-accelerated ray tracing \cite{sionna}. Although Sionna RT is a relatively new tool, it has rapidly gained traction in the literature for diverse wireless applications. For instance, authors in \cite{ns3} integrated Sionna RT with the network simulator ns-3 to establish a full-stack digital network twin. Similarly, researchers in \cite{oai_sionna} proposed a DT platform by combining OpenAirInterface (OAI) with Sionna RT to build a realistic end-to-end fifth-generation (5G) channel emulator. The sensing potential of Sionna RT was highlighted in \cite{sionna_sensing} to achieve real-time object localization through DT. In the specific context of RIS-assisted communications, researchers in \cite{sionna_ris} utilized Sionna RT to evaluate RIS cascaded channel estimation methods, including Hadamard and orthogonal matching pursuit, however, this study employed Sionna as a simulation tool rather than establishing a functional DT. 
Apart from Sionna-specific studies, researchers in \cite{ris_rt} proposed a method to derive RIS phase matrices using an RT-based DT, thereby bypassing the need for RIS channel estimation. However, this work relies exclusively on numerical simulations and does not address real-time experimental validation. To the best of our knowledge, no study in the literature utilizes Sionna-RT as a DT to optimize RIS phases without CSI acquisition for RIS channels, integrates this framework into a 3GPP-compliant emulator, and validates the performance through real-time experimental measurements.

\begin{figure*}
    \centering
    \includegraphics[width=\linewidth]{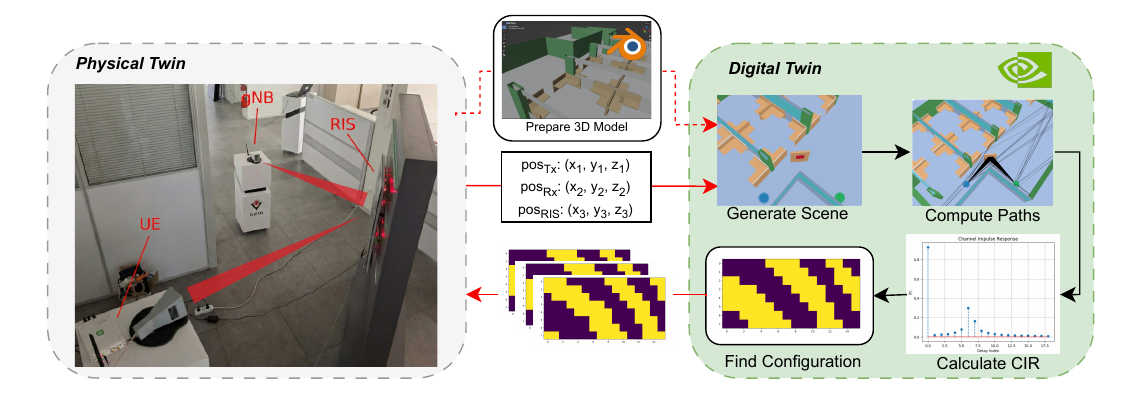}
    \caption{Proposed DT framework for RIS phase shift optimization.}
    \label{fig:framework_model}
\end{figure*}

In this paper, we propose a practical DT-driven optimization framework that utilizes RT to determine the optimal phase configurations of RIS elements, which are then directly applied to a real RIS prototype without requiring prohibitively expensive real-time channel estimation procedures. Specifically, we first construct a precise 3D model of the experimental environment using Blender and import it into the NVIDIA Sionna library. Within this digital replica, we optimize the RIS phase shifts to maximize the reference signal received power (RSRP) using two approaches, namely DT-based direct phase optimization (DT-DPO) and DT-based channel impulse response extraction (DT-CIR). To experimentally validate this technique, the phase configurations derived from the DT are applied directly to a physical RIS prototype \cite{yerliris} integrated with a USRP-based testbed running OAI \cite{oai}. Furthermore, to assess the effectiveness of our approach, we execute a conventional optimization algorithm directly in the physical environment, serving as a performance benchmark. Finally, we demonstrate that the proposed DT-based method achieves RSRP performance comparable to this physical benchmark.

The remainder of this paper is organized as follows. Section \ref{sec:system_model} introduces the system model, along with the challenges related to RIS phase optimization in real environments, and presents the proposed DT-driven RIS optimization framework and methodology. Section \ref{sec:experiments} describes the experimental setup and reports the experimental results. Finally, Section \ref{sec:conclusion} concludes the paper.

\section{Ray Tracing-based Digital Twin Framework for RIS Phase Optimization}
\label{sec:system_model}
In this section, we present the system architecture for an RIS-aided downlink communication scenario. We consider a single-input single-output system, where a base station (BS) communicates with user equipment (UE) through a 5G NR waveform. Assuming that the direct link is weak, an RIS is deployed to assist the transmission. As illustrated in Fig. \ref{fig:framework_model}, the proposed framework operates by synchronizing the physical deployment with a DT constructed using RT. In this system, the real-time coordinates of the network entities are mapped to the virtual environment to compute the optimal phase configurations, which are subsequently loaded into the physical RIS prototype.

\subsection{Signal Model}
Let $x \in \mathbb{C}$ denote the transmitted symbol on a specific subcarrier with unit power. The signal received at the UE is the sum of the signal from the direct path (if available) and the signals reflected by the RIS elements. The equivalent baseband received signal, $y \in \mathbb{C}$, can be expressed as:
\begin{equation}
   y = \left[h_d+\sum_{i = 1}^{N}h_{i}e^{j\theta_{i}}g_{i}\right]x + w,
   \label{eq:scalar_model}
\end{equation}
where $h_d=|h_d|e^{j\alpha_d}$ represents the direct channel coefficient, $N$ is the number of RIS elements with phase shift $\theta_i$, and $w \sim \mathcal{CN}(0, \sigma^2)$ is the additive white Gaussian noise. The terms $h_i=|h_i|e^{j\alpha_i}$ and $g_i=|g_i|e^{\beta_i}$ are the channel coefficients between BS and $i^{th}$ RIS element and $i^{th}$ RIS element and UE, respectively. By defining the RIS phase shift matrix as a diagonal matrix $\mathbf{\Phi} = \text{diag}(e^{j\theta_1}, \dots, e^{j\theta_N})$, the received signal in \eqref{eq:scalar_model} can be rewritten in a compact vector form as
\begin{equation}
   y= [h_d+\mathbf{g^T}\mathbf{\Phi}\mathbf{h}]x+w,
   \label{eq:rx_signal}
\end{equation}
where $\mathbf{g}=[g_1, g_2, ..., g_N]^T, \mathbf{h}=[h_1,h_2, ..., h_N]^T$ are the channel coefficient vectors between BS-RIS and RIS-UE, respectively.

Our primary objective is to maximize the received power based on RIS phase shifts as,
\begin{equation}
    \begin{split}
   &\max_{\Phi \in \mathcal{X}}\; |h_d+\mathbf{g^T}\mathbf{\Phi}\mathbf{h}|^2,\\
   s.t.&\; \theta_i \in [0,2\pi)\;\;\forall i=1,2,...,N,
   \end{split}
   \label{eq:optimization_problem}
\end{equation}
where $\mathcal{X}$ is the search space containing RIS configuration matrices. 

Ideally, to maximize the objective function in \eqref{eq:optimization_problem}, the RIS phase shifts should be adjusted such that the signals reflected from the RIS paths are coherently aligned with the signal from the direct path. The optimal phase shift for the $i$-th element is given by,
\begin{equation}
\theta_i^* = \alpha_d - (\alpha_i + \beta_i), \quad \forall i = 1, \dots, N.
\label{eq:optimal_phase}
\end{equation}

However, in this study, we consider a practical RIS architecture with 1-bit phase quantization, where each element can only take binary phase values (i.e., $\theta_i \in \{0, \pi\}$). To accommodate this hardware constraint, the optimal continuous phases in \eqref{eq:optimal_phase} are mapped to the nearest discrete level based on the minimum Euclidean distance criterion:
\begin{equation}
\bar{\theta}_i = \underset{\theta \in \{0, \pi\}}{\arg \min} | e^{j\theta_i^*} - e^{j\theta} |.
\label{eq:quantized_phase}
\end{equation}

\subsection{Challenges in Real-Time RIS Phase Optimization}
The optimization problem described in \eqref{eq:optimization_problem} under discrete phase constraint requires selecting the optimal phase configuration matrix $\mathbf{\Phi}$ from a huge search space $\mathcal{X}$. For an RIS with $N$ elements and 1-bit quantization, there exist $|\mathcal{X}|=2^N$ possible configurations. Consequently, identifying the global optimum through an exhaustive search incurs excessive computational latency. Although the existing methods designed to find local optima reduce the search space, they still impose significant timing overheads, which are unsuitable for mobile scenarios. Therefore, a more efficient approach to reducing the search space is essential to enable real-time RIS optimization.

Furthermore, while the analytical solution in \eqref{eq:optimal_phase} appears straightforward, its practical implementation is challenging. As indicated by the equation, calculating $\theta_i^*$ requires precise knowledge of the channel phase components for each RIS element. In a conventional setup, acquiring this instantaneous CSI for passive RIS elements requires complex pilot transmission schemes, where the number of pilots scales with $N$. For a large RIS, this results in significant signaling overhead, thereby severely degrading spectral efficiency.

\subsection{Proposed Ray Tracing-Based Digital Twin Framework}\label{sec:proposed_method}
To address the above mentioned challenges, we propose a geometry-based DT approach that computes the necessary phase configurations without relying on real-time over-the-air training. This approach avoids the time-consuming sequential physical measurement process in the physical world by enabling parallelized virtual evaluations in the DT, thereby significantly accelerating the optimization. The core operation of the proposed framework relies on mapping the real-time physical coordinates of the BS and the UE onto the DT setting, which is constructed upon a 3D map of the environment. Once this spatial configuration is established, the DT utilizes ray tracing to deterministically compute the electromagnetic interactions, as illustrated in Fig. \ref{fig:framework_model}. The DT transmits only a few optimized configuration matrices to the \textit{Physical Twin} to minimize the search space of the conventional problem defined in \eqref{eq:optimization_problem}. Within this framework, the RIS optimization is carried out through two approaches, both of which are experimentally validated to assess the performance of the DT-based RIS phase optimization.

\subsubsection{DT-based Direct Phase Optimization (DT-DPO)}

In this approach, the RIS phase optimization is performed entirely within the DT. The ray tracer computes the received signal power for varying RIS phase configurations and employs one of the RIS optimization algorithms within the DT to maximize the objective function defined in \eqref{eq:optimization_problem}. The output is the search space with the optimized phase configuration matrices, denoted as $\tilde{\mathcal{X}}_{DT-DPO}$, which represent the candidate configurations that have high probability of maximizing the objective function in \eqref{eq:optimization_problem}. These matrices, where $|\tilde{\mathcal{X}}_{DT-DPO}| \ll |\mathcal{X}|=2^N$, are then directly applied to the physical RIS controller. Evaluating \eqref{eq:optimization_problem} over this reduced subset yields a suboptimal yet highly effective solution in a fraction of the time required for an exhaustive physical search.

\subsubsection{DT-based Channel Impulse Response Extraction (DT-CIR)} 
Alternatively, the DT can be utilized as a deterministic channel estimator. Instead of outputting the final phase matrix, the system extracts the CIR using DT platform's propagation solver. Specifically in Sionna, the propagation paths are first determined by executing the \texttt{{scene.compute\_paths()}} function. Subsequently, the CIR is generated via \texttt{{paths.cir()}}. This provides the channel components, specifically the direct channel ($h_d$) and the cascaded channel ($\mathbf{g^T}\mathbf{h}$). Having obtained these channel parameters virtually, the optimal phase shifts are calculated analytically using \eqref{eq:optimal_phase} and \eqref{eq:quantized_phase}. This mode offers flexibility, allowing for the application of different quantization or beamforming strategies on the extracted channel data without re-running the ray-tracing simulation. 

\begin{figure}[t]
\centering
\begin{tikzpicture}[
    auto,
    scale=0.85, 
    transform shape,
    block/.style = {rectangle, draw, fill=blue!10, text width=2.4cm, text centered, rounded corners, minimum height=1.1cm, font=\scriptsize},
    line/.style = {draw, -latex', thick, rounded corners},
    dashedbox/.style = {draw, dashed, inner sep=10pt, rounded corners, fill=gray!5, font=\scriptsize, align=center}
]

    \node [block, fill=orange!15] (blender) {3D Modeling \\ (Blender)};
    \node [block, right=0.6cm of blender] (sionna) {Import \& Setup \\ \texttt{load\_scene()}};
    \node [block, right=0.6cm of sionna] (virt_pos) {Virtual Positioning \\ (Tx, Rx, RIS)};
    
    \node [block, below=0.8cm of virt_pos, text width=3.2cm] (opt) {DT Optimization \\ \textbf{Approach 1:} DT-DPO \\ \textbf{Approach 2:} DT-CIR};
    \node [block, left=1.0cm of opt, fill=yellow!20] (save) {Save Optimized Configuration Matrices \\ ($\tilde{\mathcal{X}}_{DT}$) or CIR};

    \node [block, below=1.9cm of save, xshift=0.5cm, fill=cyan!15] (baseline) {Measure RSRP \\ (All-Zero Configuration)};
    
    \node [block, right=0.8cm of baseline, fill=green!10] (phy_setup) {Physically Matching the Coordinates of Tx, Rx and RIS};

    \node [block, below=1.2cm of baseline, xshift=-2.2cm] (load_dt) {Load Configuration Matrices into RIS Controller};
    \node [block, below=0.6cm of load_dt, fill=cyan!15] (meas_dt) {Measure RSRP Maximizing \eqref{eq:optimization_problem} (DT Method)};
    
    \node [block, below=1.2cm of baseline, xshift=2.2cm] (bench_alg) {Run Physical \\ Iterative Search Method};
    \node [block, below=0.6cm of bench_alg, fill=cyan!15] (meas_bench) {Measure RSRP (Benchmark)};

    \path [line] (blender) -- (sionna);
    \path [line] (sionna) -- (virt_pos);
    \path [line] (virt_pos) -- (opt);
    \path [line] (opt) -- (save);
    
    \path [line] (phy_setup) -- (baseline);
    
    \coordinate (split_point) at ($(baseline.south) + (0,-0.6)$);
    \path [line] (baseline.south) -- (split_point);
    \path [line] (split_point) -| (load_dt.north);
    \path [line] (split_point) -| (bench_alg.north);
    
    \path [line] (load_dt) -- (meas_dt);
    \path [line] (bench_alg) -- (meas_bench);
    
    \draw [line, dashed, color=red] (save.west) -- ++(-2.0,0) |- node [pos=0.25, left, font=\tiny, color=black, align=right] {} (load_dt.west);

    \draw [line, dashed, color=red] (virt_pos.east) -- ++(0.6,0) |- node [pos=0.25, right, font=\tiny, color=black, align=left] {} (phy_setup.east);

    \begin{scope}[on background layer]
        \node [dashedbox, fit=(blender) (sionna) (virt_pos) (opt) (save), label=above:\textbf{Digital Twin Domain}] (dt_layer) {};
        
        \node [dashedbox, fit=(phy_setup) (baseline) (load_dt) (meas_dt) (bench_alg) (meas_bench), label=above:\textbf{Physical Domain}] (phy_layer) {};
    \end{scope}

\end{tikzpicture}
\caption{Flowchart of the experimental methodology.}
\label{fig:exp_flowchart}
\end{figure}
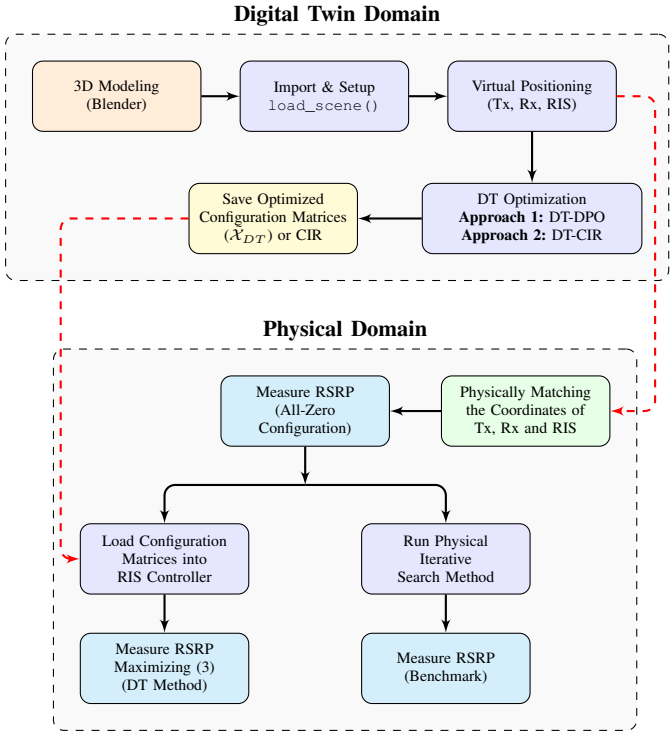

\section{Experimental Setup and Performance Evaluation} \label{sec:experiments}

In this section, we present the experimental setup to validate the proposed DT-driven optimization framework and discuss the measurement results from a real-world indoor scenario.

\begin{figure}[t]
    \centering
    \includegraphics[width=0.465\linewidth]{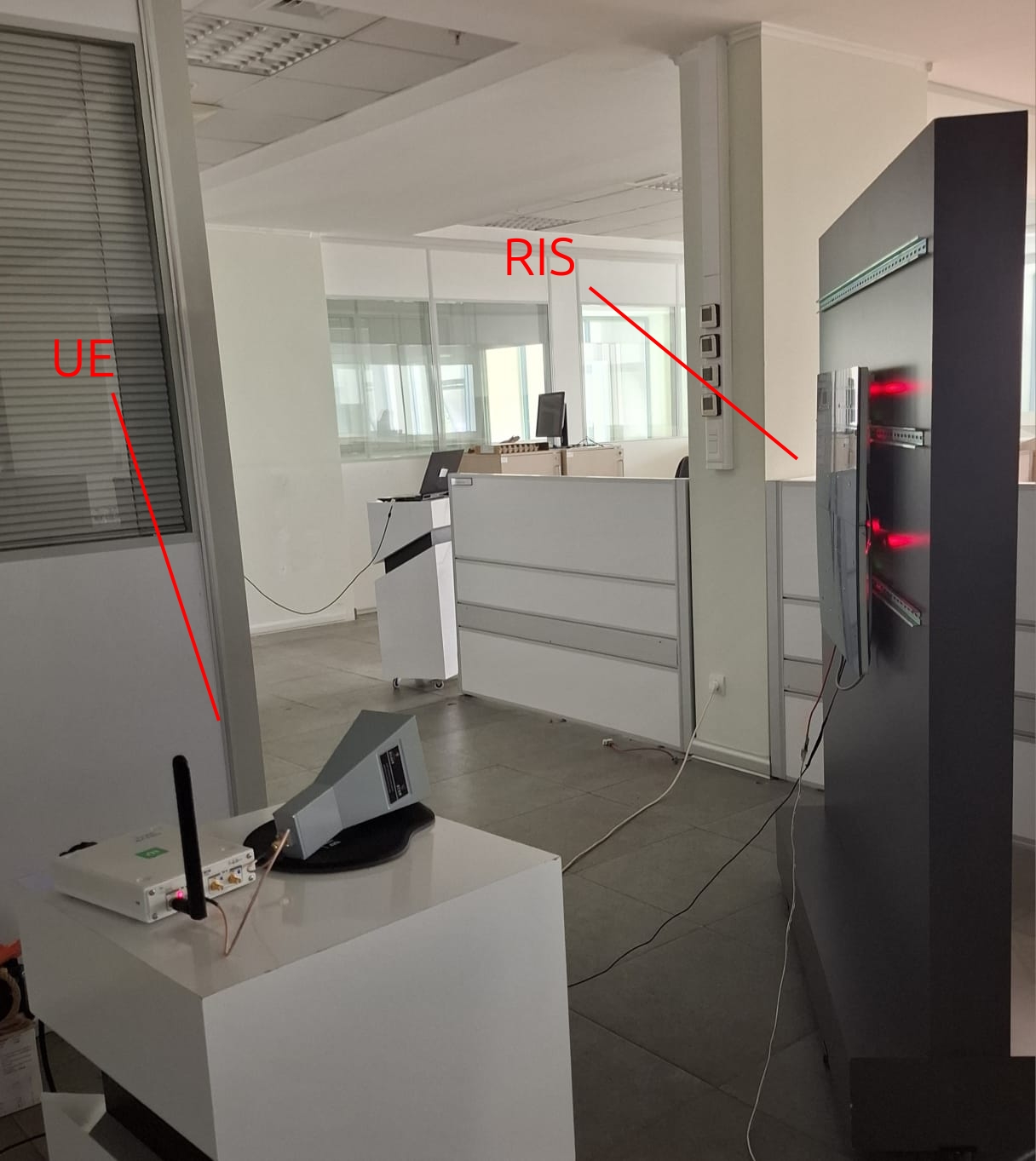}
    \hfill
    \includegraphics[width=0.52\linewidth]{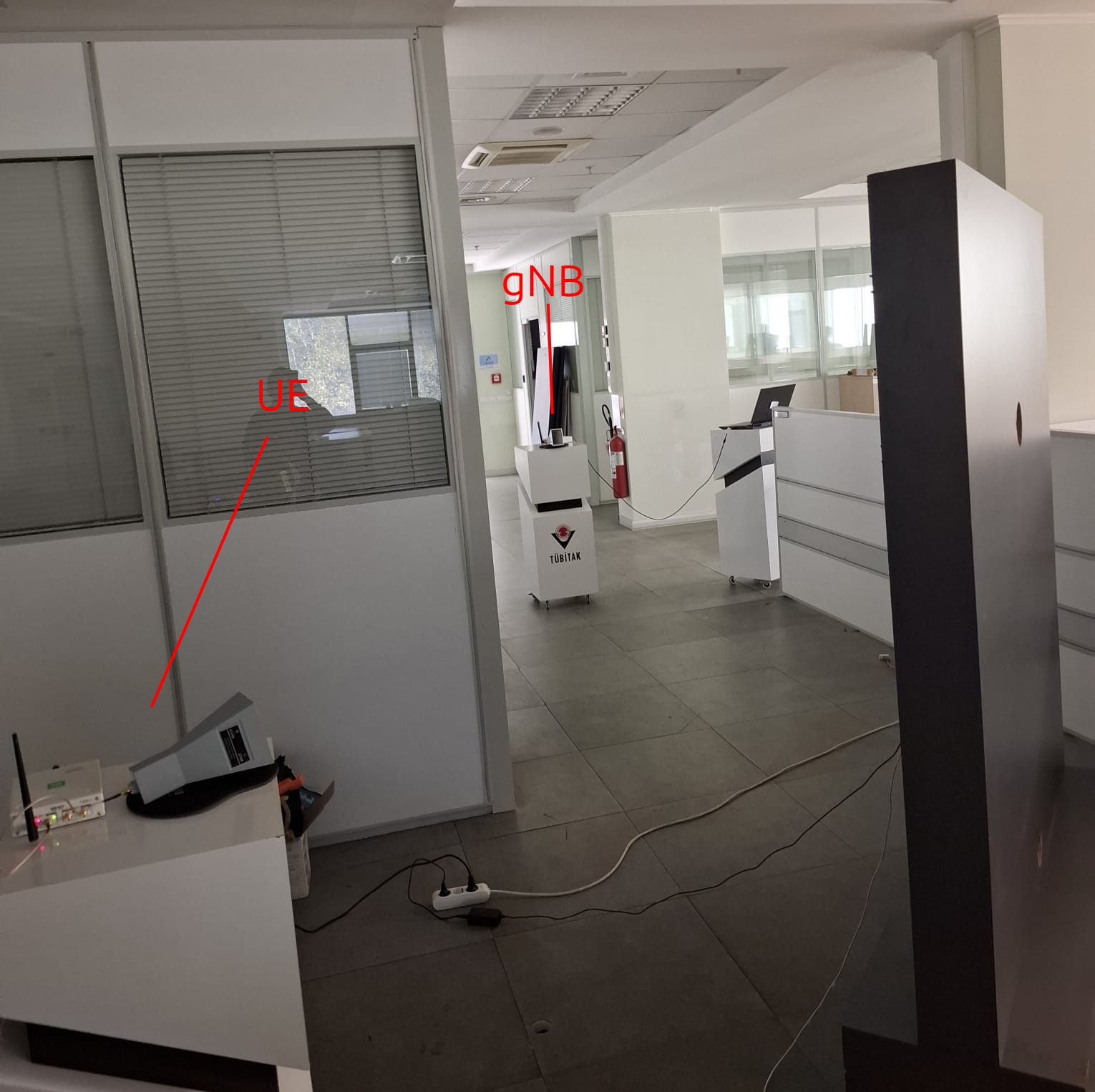}
    \caption{Experimental setup.}
    \label{fig:exp_setup}
\end{figure}

\subsection{Experimental Environment and Hardware Setup} 
The measurements were conducted in an indoor office environment comprising concrete walls, office partitions, and tables, as depicted in Fig. \ref{fig:exp_setup}. The hardware setup employs a software defined radio (SDR) based testbed. We utilized two USRP B210 platforms to serve as the BS (transmitter) and the UE (receiver). To emulate the 5G physical layer stack, the OAI platform was deployed on the host computers. The system operates in the FR1 n78 band at a center frequency of 3.62 GHz.

To maximize the gain of the RIS-assisted downlink channel, both the transmitter and receiver RF chains are equipped with directional horn antennas aligned towards the RIS. Secondary omni-directional antennas are connected to both USRPs to maintain the uplink control signaling and synchronization required by the OAI 5G emulator.

For the RIS, we utilized a 1-bit RIS prototype, operating in FR1 n78 frequency bands \cite{yerliris}. Specifically, the RIS utilized in the experimental setup consists of an $8\times16$ array of unit cells (128 elements), and its configuration is controlled by the UE host via a USB interface. The key experimental parameters are summarized in Table \ref{tab:exp_params}.

\begin{table}[ht!]
\centering
\caption{Experimental Configurations}
\renewcommand{\arraystretch}{1.2} 
\setlength{\tabcolsep}{5pt}
\label{tab:exp_params}
\begin{tabular}{lc}
\hline
\multicolumn{1}{c}{\textbf{Parameter}} & \textbf{Value} \\ \hline
\multicolumn{1}{c|}{Emulator Platform} & OpenAirInterface5G \\
\multicolumn{1}{c|}{Carrier Frequency} & 3.62 GHz (n78) \\
\multicolumn{1}{c|}{SDR Hardware} & 2 $\times$ USRP B210 \\
\multicolumn{1}{c|}{Downlink Antennas} & Directional Horn \\
\multicolumn{1}{c|}{Uplink/Control Antennas} & Omnidirectional \\
\multicolumn{1}{c|}{RIS Structure} & 1-bit, $8 \times 16$ elements \\
\multicolumn{1}{c|}{Waveform} & 5G NR \\
\multicolumn{1}{c|}{Tx-RIS Distance} & 420 cm \\
\multicolumn{1}{c|}{RIS-Rx Distances} & \{183.7, 185.7, 418.5, 419.4\} cm \\
\hline
\end{tabular}
\end{table}

\subsection{Digital Twin Construction and Methodology} 
To realize the proposed framework, a precise 3D model of the measurement environment was first created using the Blender software, replicating the physical geometry of the walls, partitions, and desks etc. This 3D mesh was then imported into the RT engine of Sionna (v19.0.2) using the \texttt{load\_scene("scene.xml")} function to establish the digital replica.

\begin{figure}[t]
    \centering
    \includegraphics[width=0.49\linewidth]{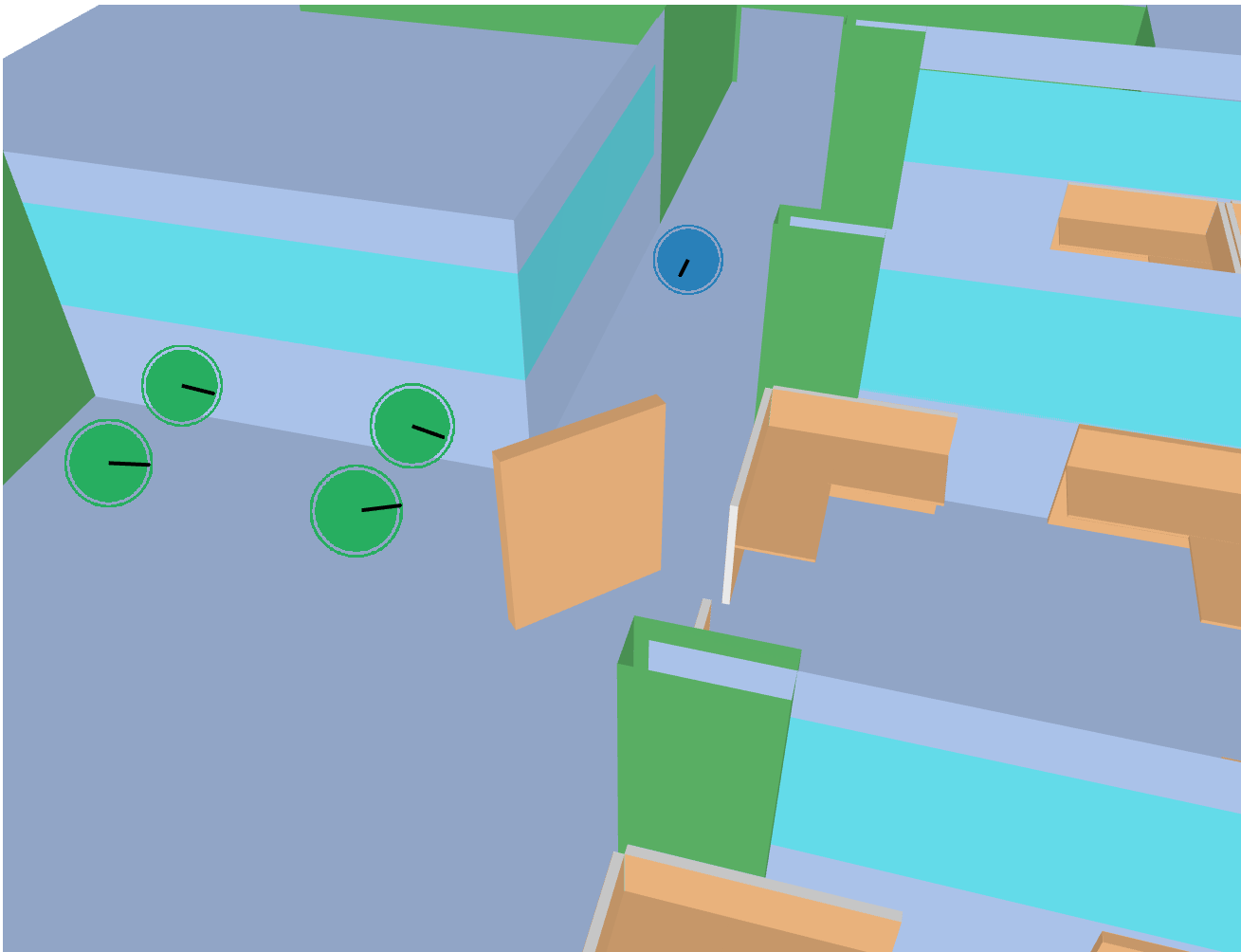}
    \hfill
    \includegraphics[width=0.49\linewidth]{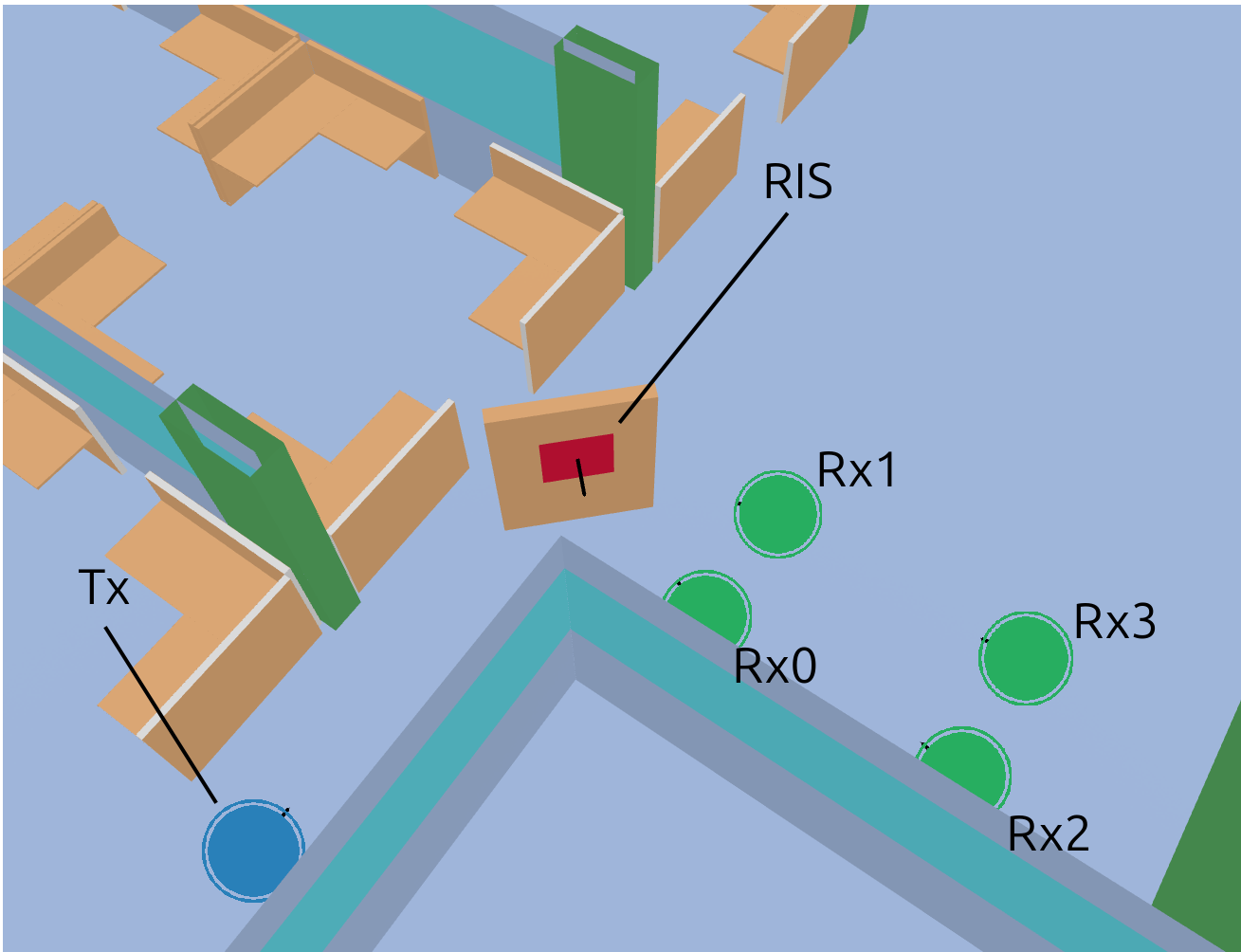}
    \caption{3D Model of the environment viewed in Sionna.}
    \label{fig:sionna_view}
\end{figure}

\begin{figure}[t]
    \centering
    \begin{minipage}{0.32\linewidth} \centering \small \textbf{Physical Opt.} \end{minipage}
    \hfill
    \begin{minipage}{0.32\linewidth} \centering \small \textbf{DT-DPO} \end{minipage}
    \hfill
    \begin{minipage}{0.32\linewidth} \centering \small \textbf{DT-CIR} \end{minipage}
    \par\medskip 
    \begin{subfigure}{\linewidth}
        \centering
        \includegraphics[width=0.32\linewidth]{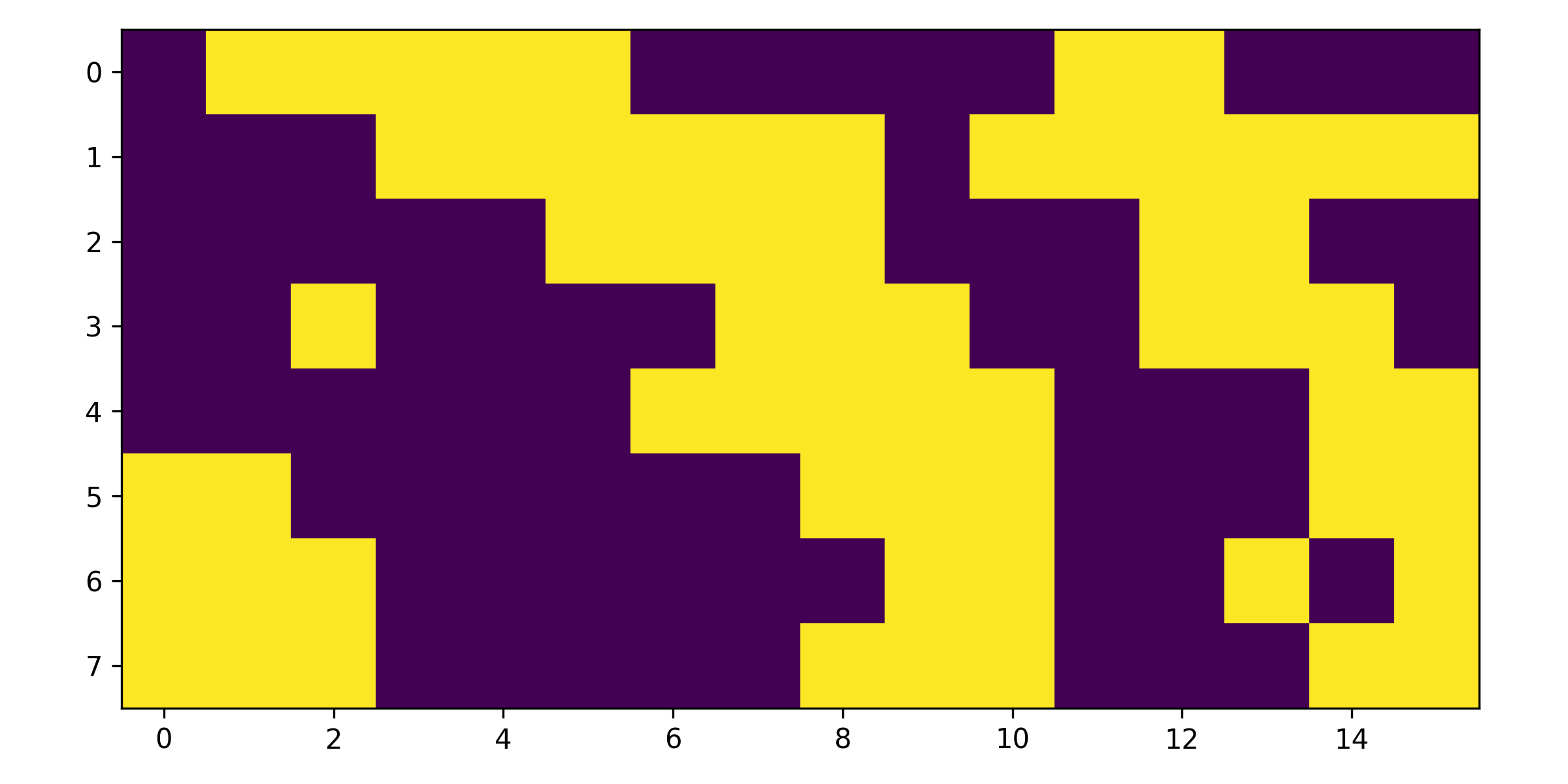} \hfill
        \includegraphics[width=0.32\linewidth]{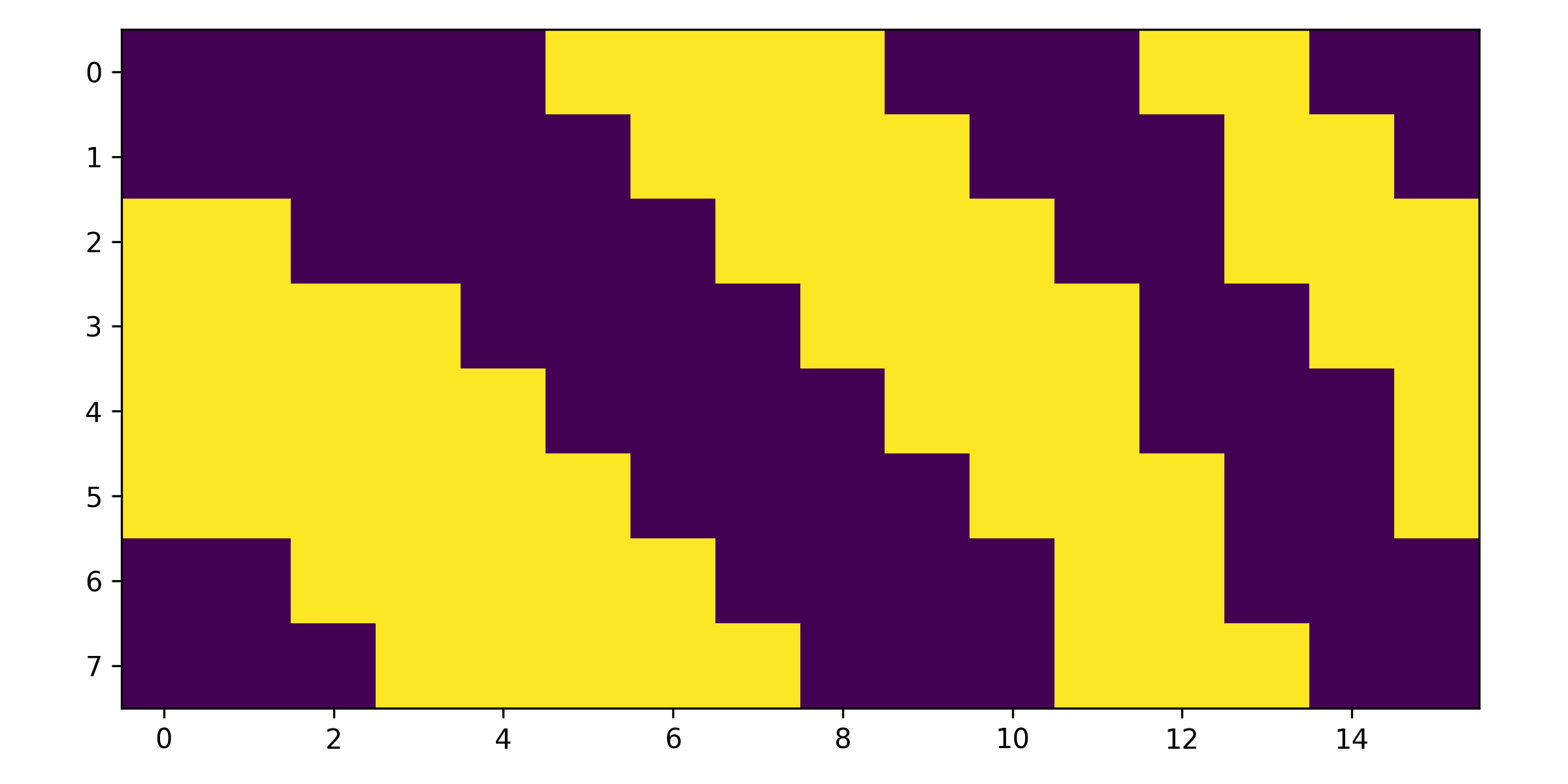}  \hfill
        \includegraphics[width=0.32\linewidth]{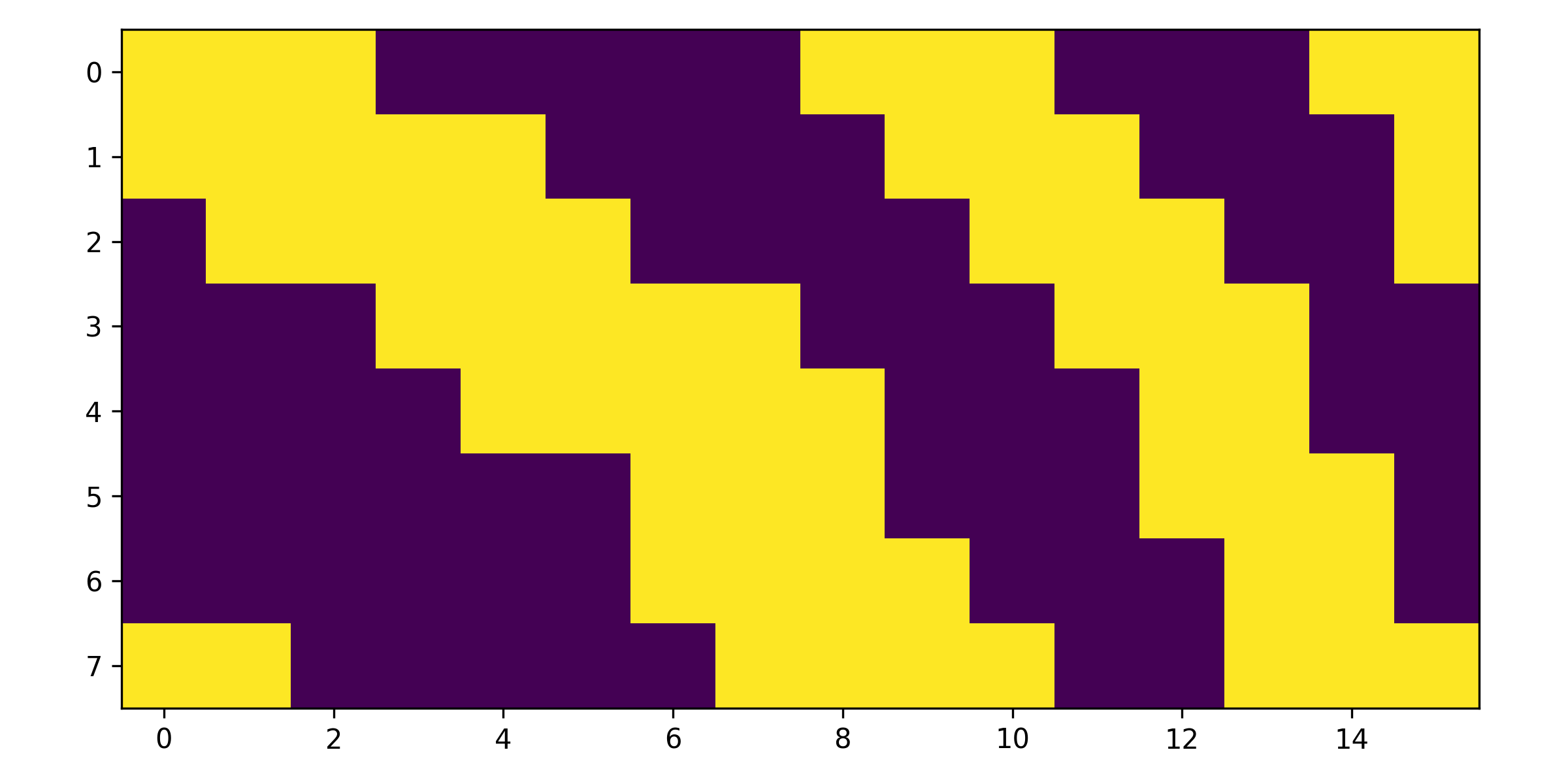}
        \caption{Rx0} 
        \label{subfig:rx0}
    \end{subfigure}
    \par\medskip

    \begin{subfigure}{\linewidth}
        \centering
        \includegraphics[width=0.32\linewidth]{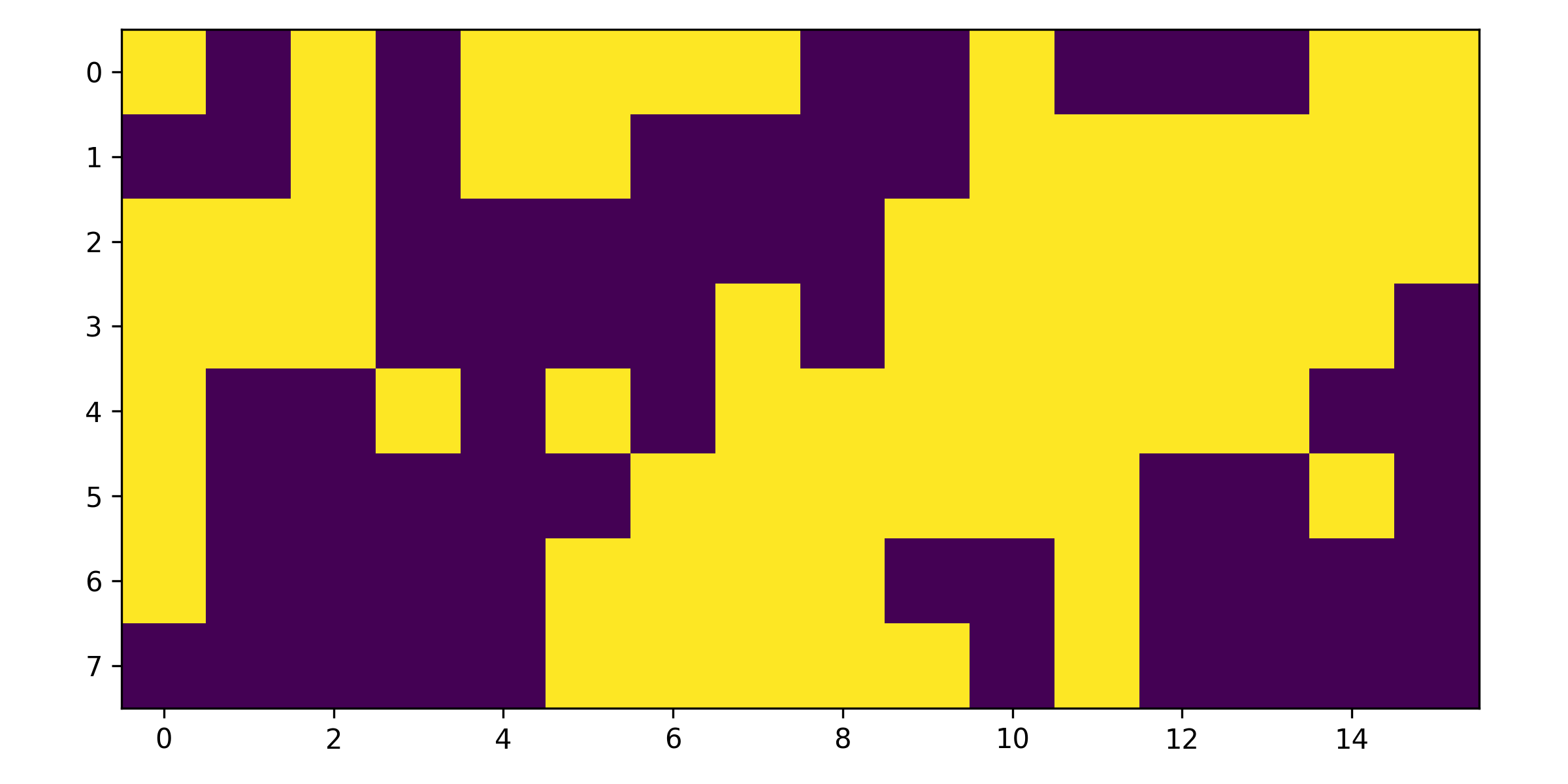} \hfill
        \includegraphics[width=0.32\linewidth]{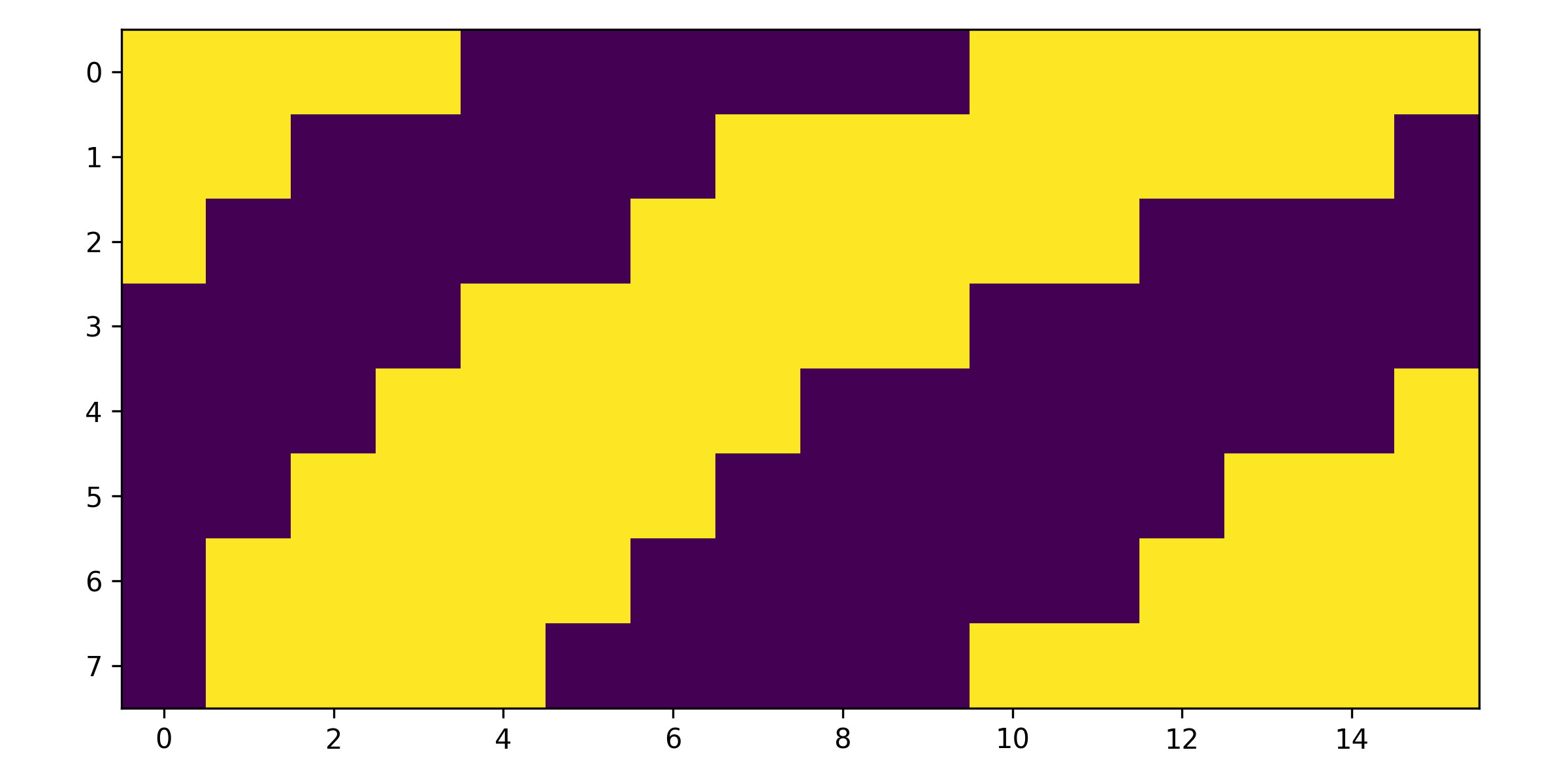} \hfill
        \includegraphics[width=0.32\linewidth]{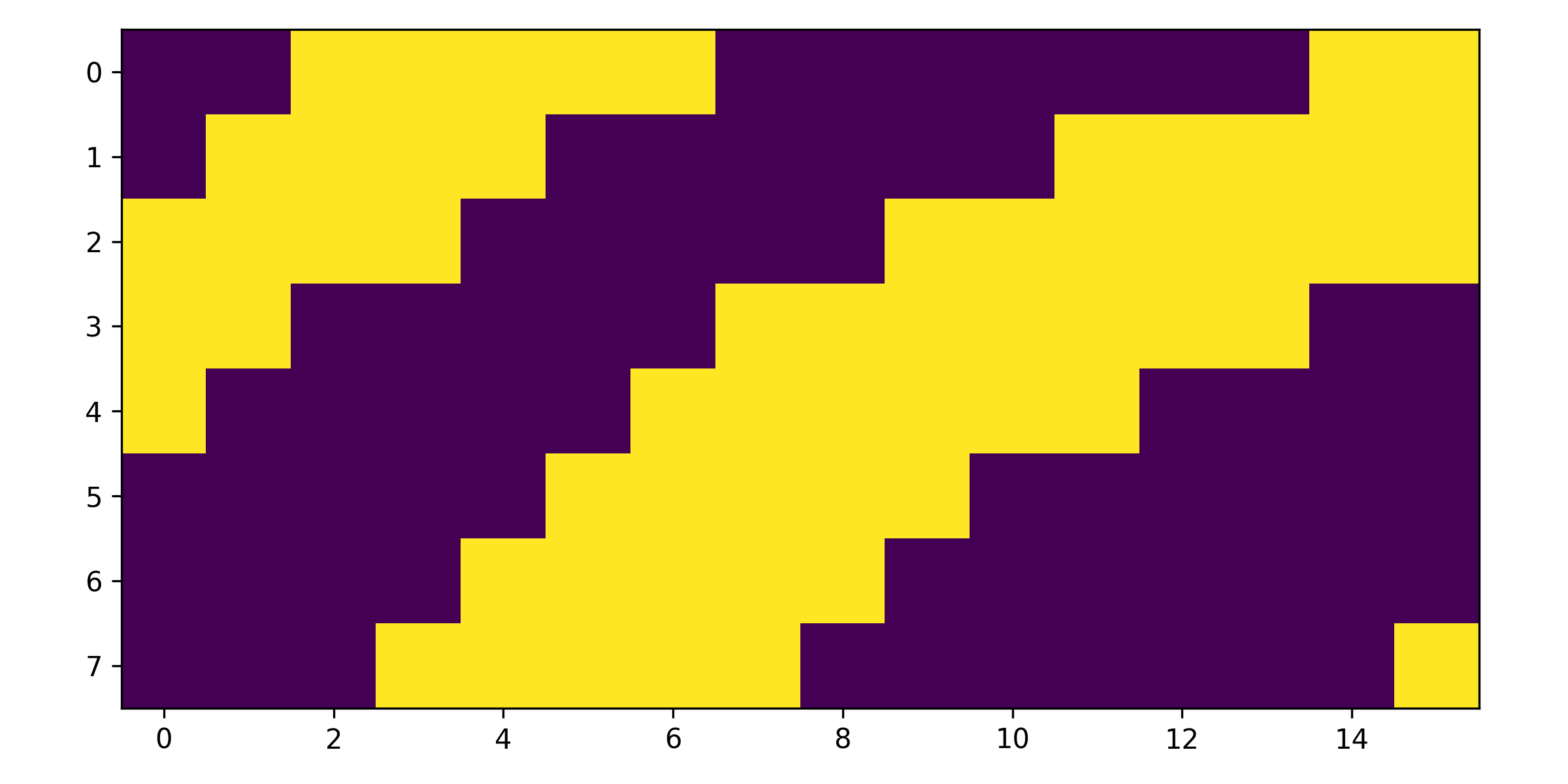}
        \caption{Rx1} 
        \label{subfig:rx1}
    \end{subfigure}
    \par\medskip

    \begin{subfigure}{\linewidth}
        \centering
        \includegraphics[width=0.32\linewidth]{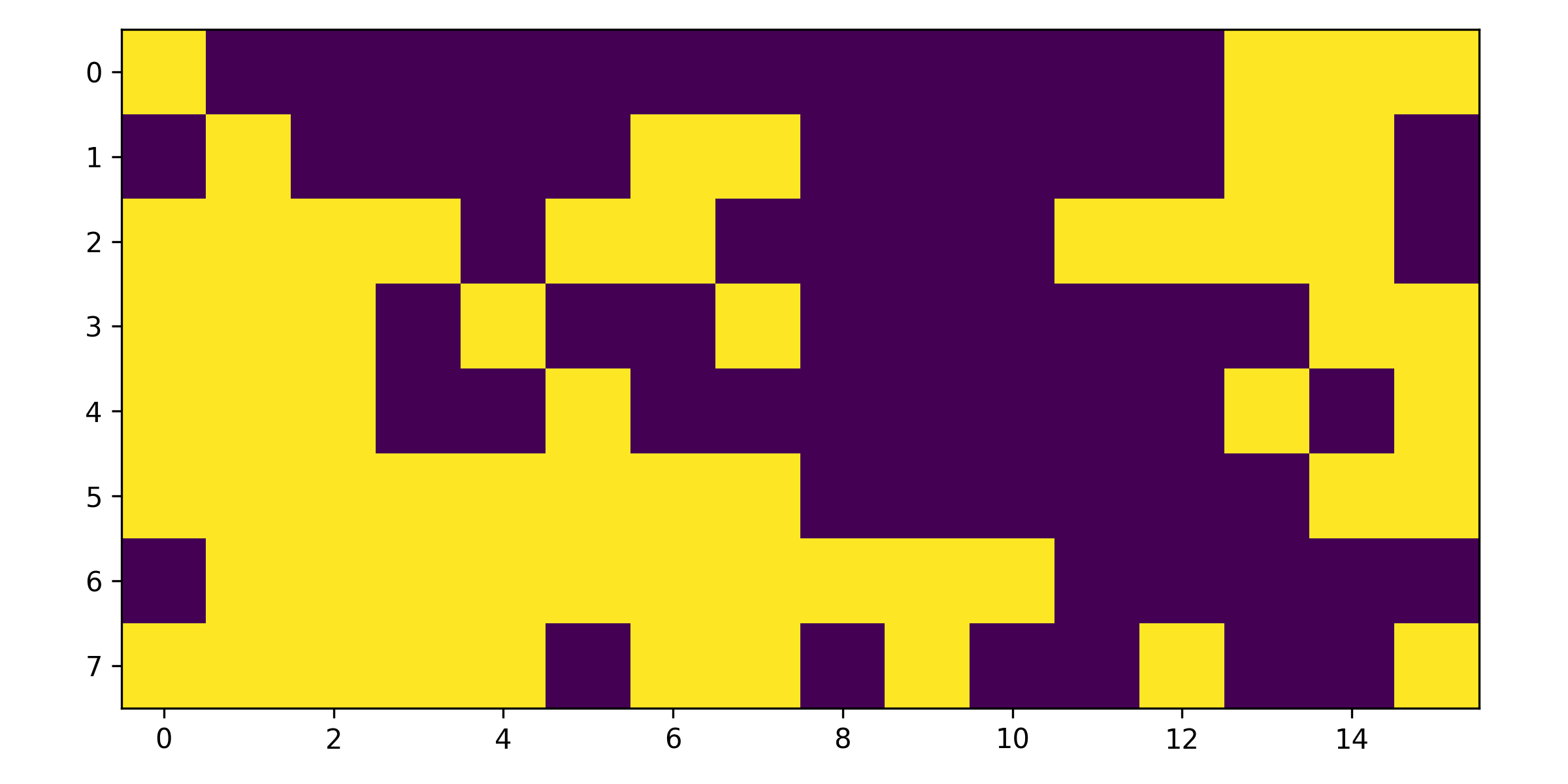} \hfill
        \includegraphics[width=0.32\linewidth]{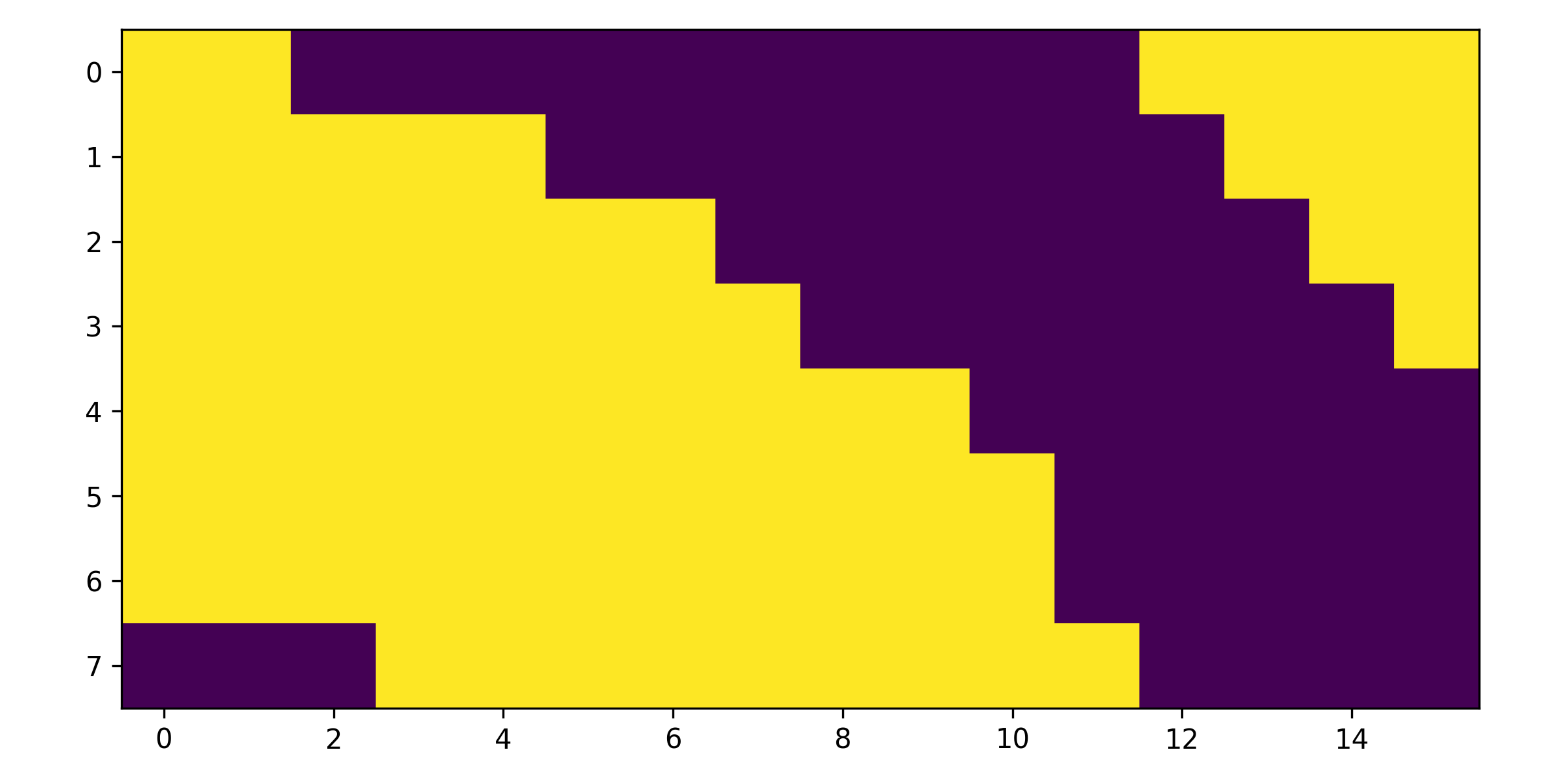} \hfill
        \includegraphics[width=0.32\linewidth]{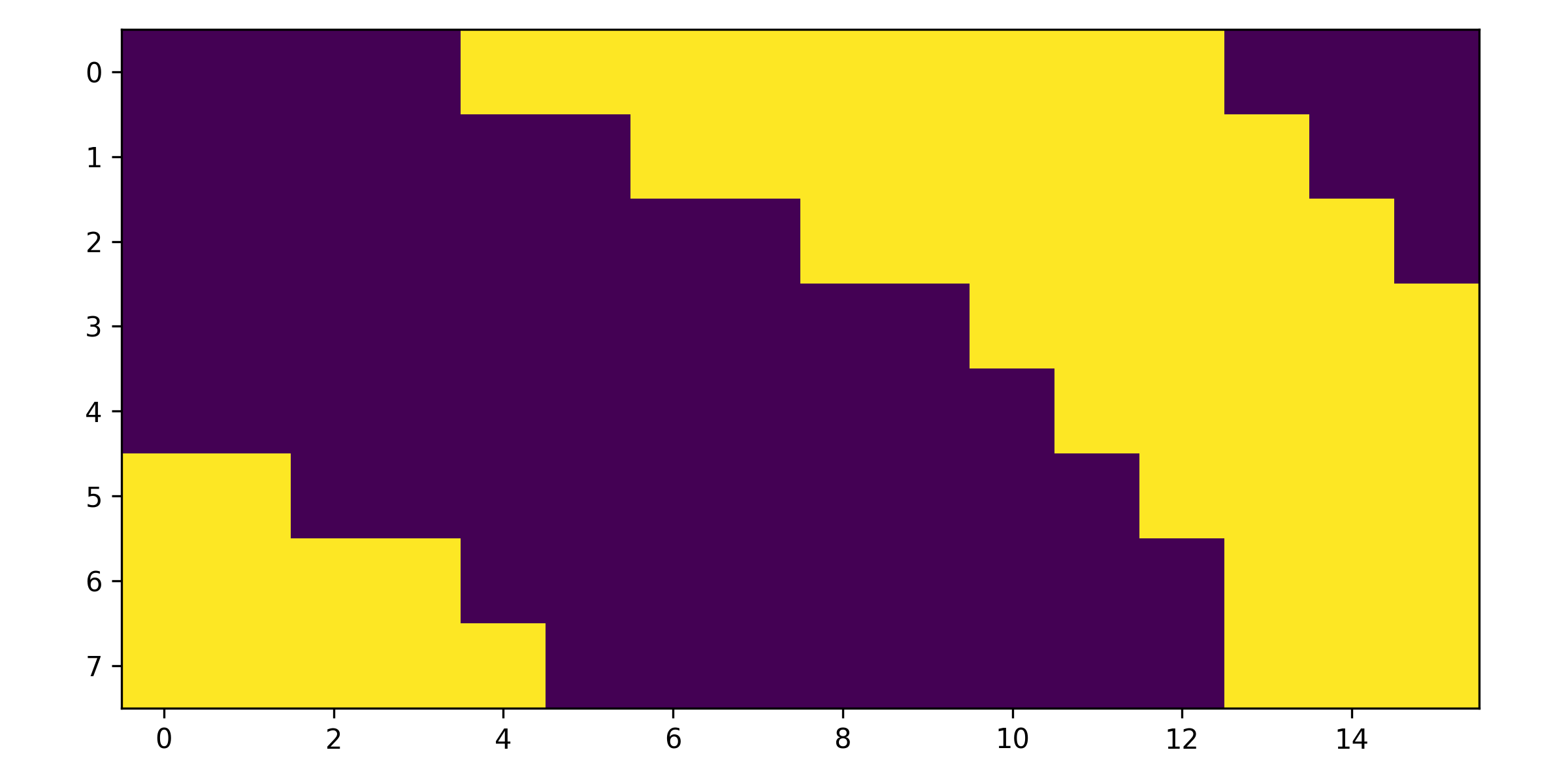}
        \caption{Rx2} 
        \label{subfig:rx2}
    \end{subfigure}\par\medskip
    \begin{subfigure}{\linewidth}
        \centering
        \includegraphics[width=0.32\linewidth]{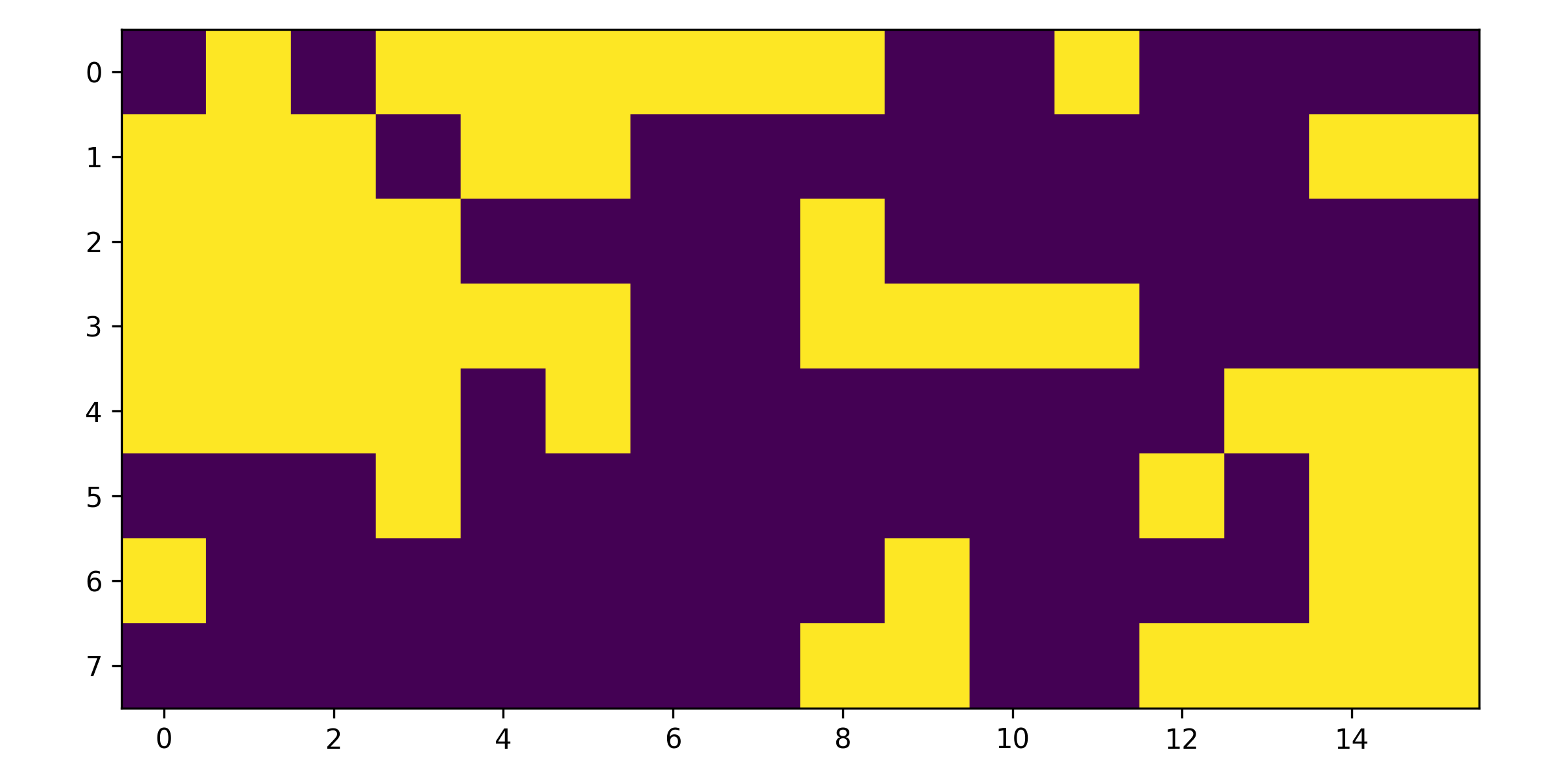} \hfill
        \includegraphics[width=0.32\linewidth]{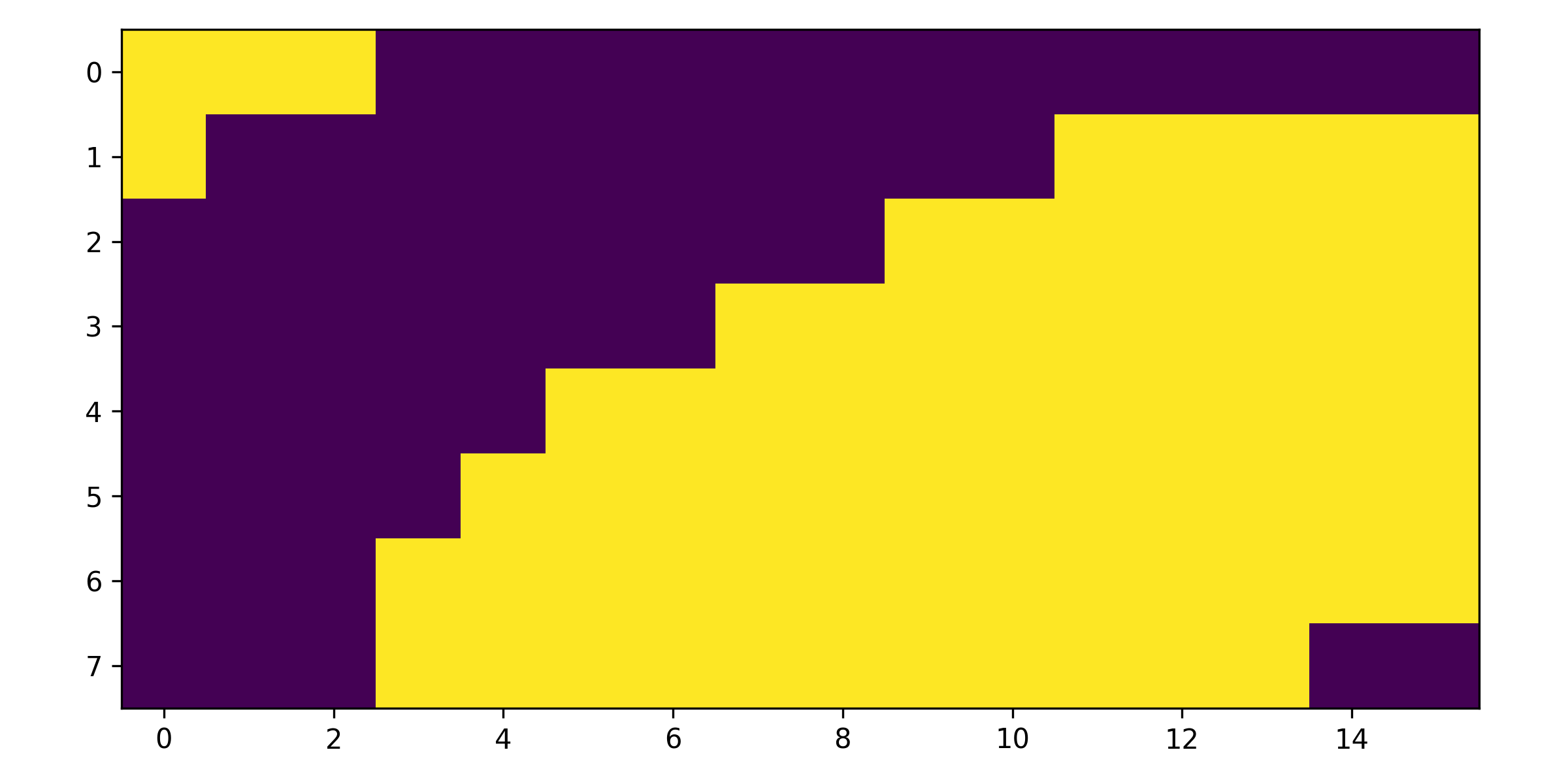} \hfill
        \includegraphics[width=0.32\linewidth]{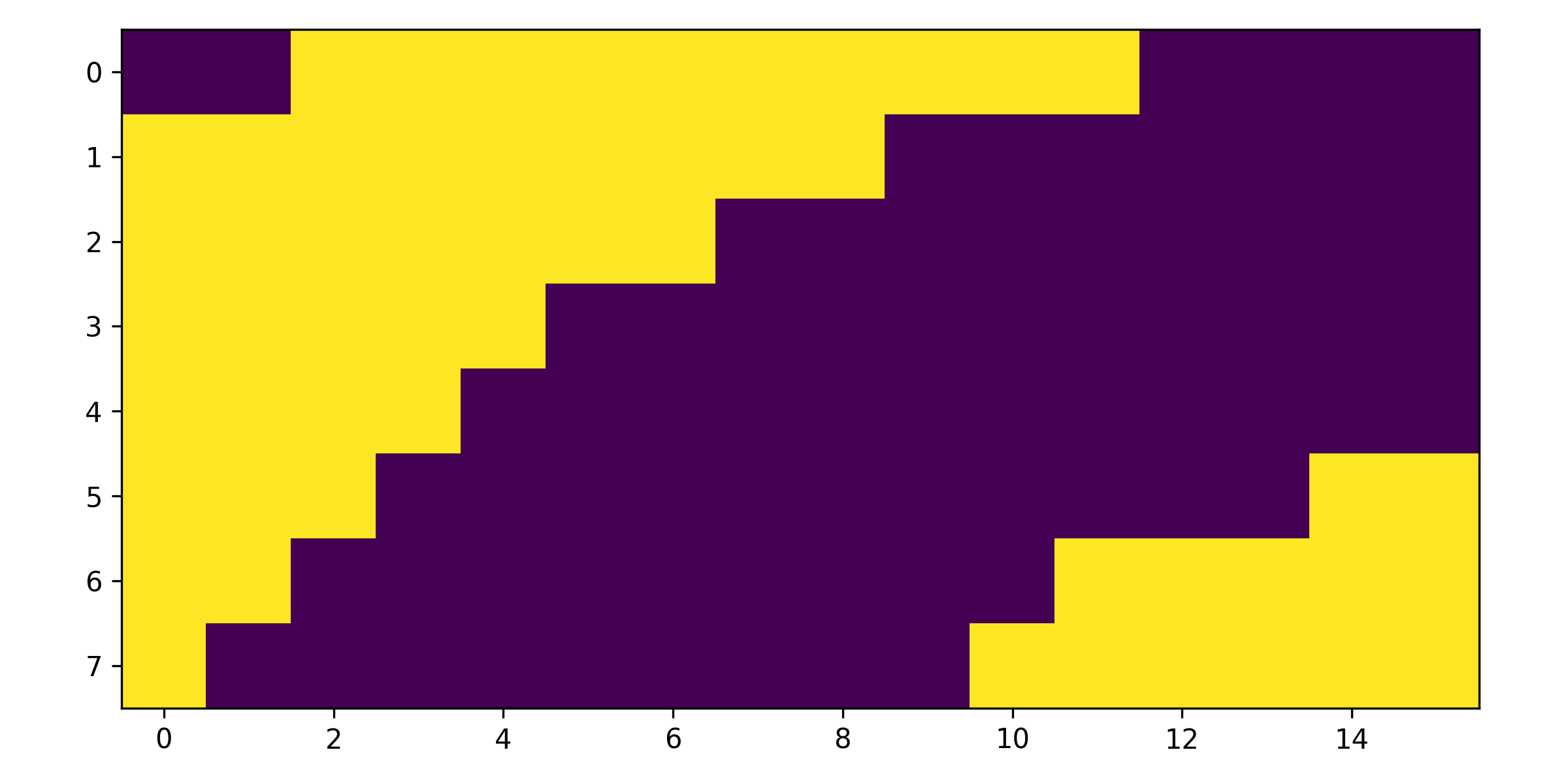}
        \caption{Rx3} 
        \label{subfig:rx3}
    \end{subfigure}

    \caption{RIS phase configurations for different UE locations (purple: OFF state, yellow: ON state).}
    \label{fig:all_patterns}
\end{figure}

Within the Sionna environment, the simulation scenario was aligned with the physical setup by positioning the virtual Tx, RIS, and four distinct Rx locations (Rx0-Rx3) at the precise coordinates selected for the experimental validation, as shown in Fig. \ref{fig:sionna_view}. We perform optimization using the two approaches defined in Section \ref{sec:proposed_method}: \begin{enumerate} \item DT-DPO: An iterative algorithm \cite{iterative} was executed within the DT to maximize the RSRP. To ensure robustness against phase ambiguities, the DT reduced the vast search space $\mathcal{X}$ to a minimal candidate set $\tilde{\mathcal{X}}_{DT-DPO}$ ($|\tilde{\mathcal{X}}_{DT-DPO}|=2$), containing only the configuration matrix optimized with iterative algorithm and its binary inverse, which were subsequently transmitted to the physical twin. \item DT-CIR: The ideal channel phases were extracted from Sionna. These continuous phase values were then quantized to 1-bit precision and mapped to the corresponding bit-inverted binary matrix required by the physical RIS hardware. \end{enumerate}

The optimized RIS configurations obtained from the DT for each UE location were recorded and subsequently loaded onto the physical RIS prototype. Finally, the RSRPs were measured in the physical setup to evaluate the efficacy of the DT-based configurations compared to the baseline scenarios.

\subsection{Measurement Results} To evaluate the effectiveness of the proposed DT-based framework, we compare the RSRP gain achieved by the DT-generated configurations against a "Physical Benchmark." The benchmark corresponds to the result of iterative search method performed directly on the physical setup, representing a local optimum performance for the given specific location. The RSRP gain is defined as the power improvement relative to the all-zero case, where all RIS elements are configured off.

\begin{figure}[t]

    \centering
    \begin{subfigure}{\linewidth}
        \centering
    \includegraphics[width=0.7\linewidth]{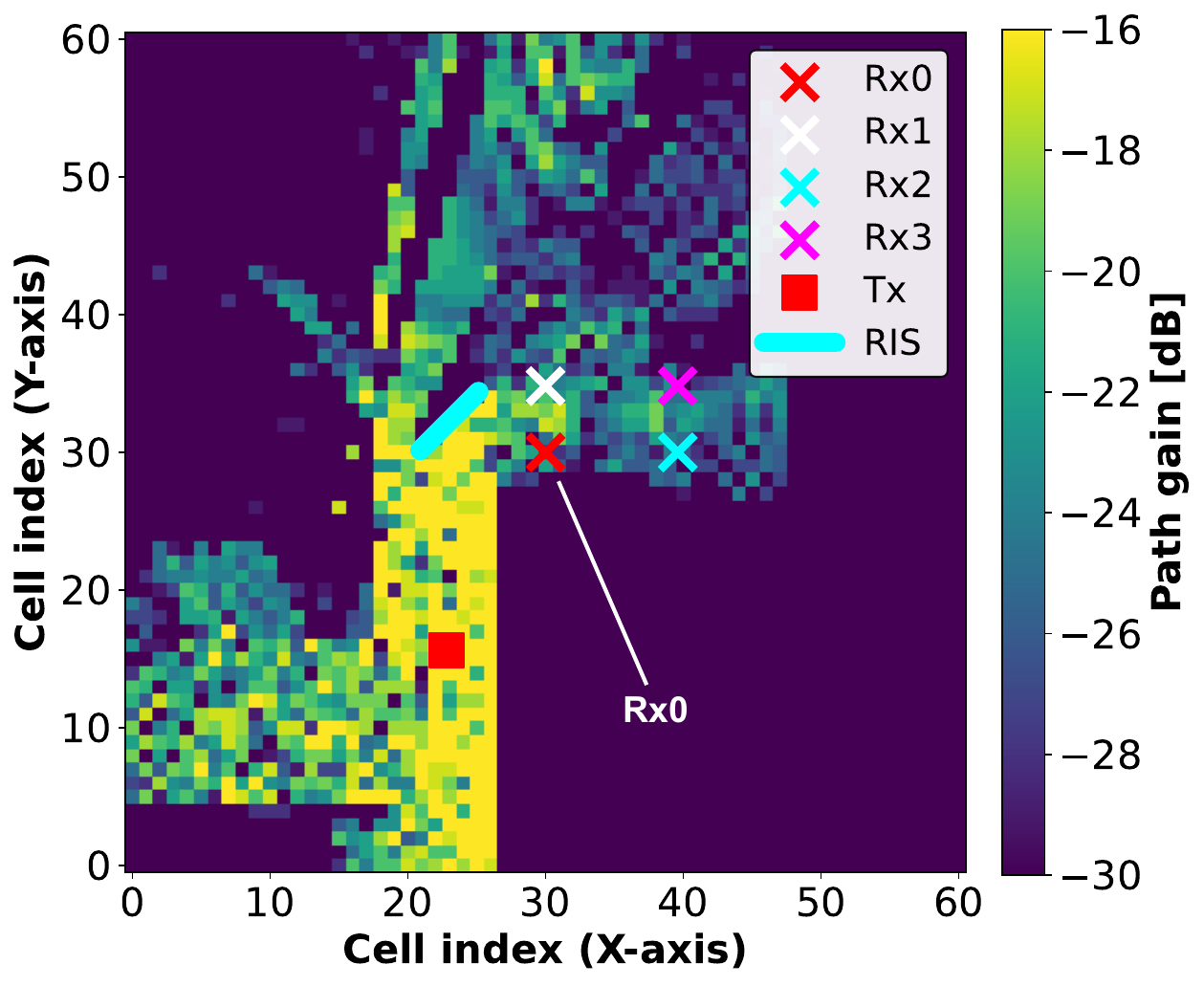}
    \caption{Scenario with the unconfigured RIS (all-zero phases).}
    \end{subfigure}
    \hfill
    \begin{subfigure}{\linewidth}
        \centering
    \includegraphics[width=0.7\linewidth]{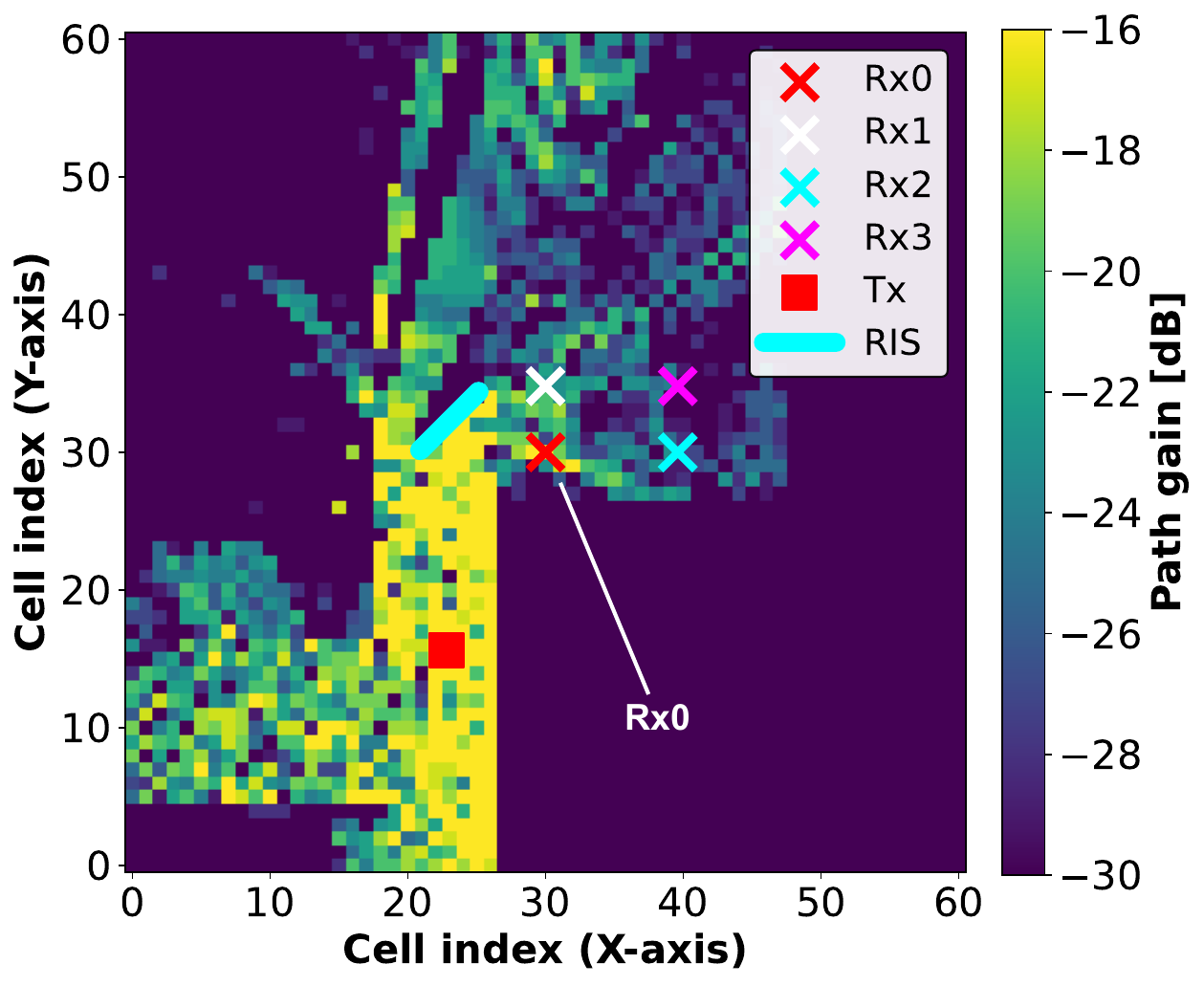}
    \caption{Scenario with the DT-DPO configuration shown in Fig. 5(a).}
    \end{subfigure}
    \caption{Simulated coverage maps.}
    \label{fig:coverage_maps}
\end{figure}

\begin{figure}[t]
\centering
\begin{tikzpicture}
    \begin{axis}[
        ybar,
        bar width=8pt,
        width=0.95\columnwidth,
        height=6cm,
        symbolic x coords={Rx0, Rx1, Rx2, Rx3},
        xtick=data,
        ylabel={RSRP Gain (dB)},
        xlabel={Receiver Locations},
        ymin=-2, ymax=20,
        nodes near coords,
        nodes near coords style={font=\footnotesize, rotate=90, anchor=west},
        legend style={at={(0.5,1.15)}, anchor=south, legend columns=-1},
        ymajorgrids=true,
        grid style=dashed,
    ]
        \addplot[style={fill=green!60!black, mark=none}] coordinates {(Rx0,16) (Rx1,8) (Rx2,12.5) (Rx3,4.5)};
        \addplot[style={fill=blue!60, mark=none}] coordinates {(Rx0,12) (Rx1,4) (Rx2,9.5) (Rx3,2.8)}; 
        \addplot[style={fill=orange!80, mark=none}] coordinates {(Rx0,9.5) (Rx1,4) (Rx2,2.5) (Rx3,-0.5)};
        \legend{Baseline, DT-DPO, DT-CIR}
    \end{axis}
\end{tikzpicture}
\caption{Comparison of RSRP gains achieved by Physical Benchmark and DT-based methods.}
\label{fig:rsrp_results}
\end{figure}
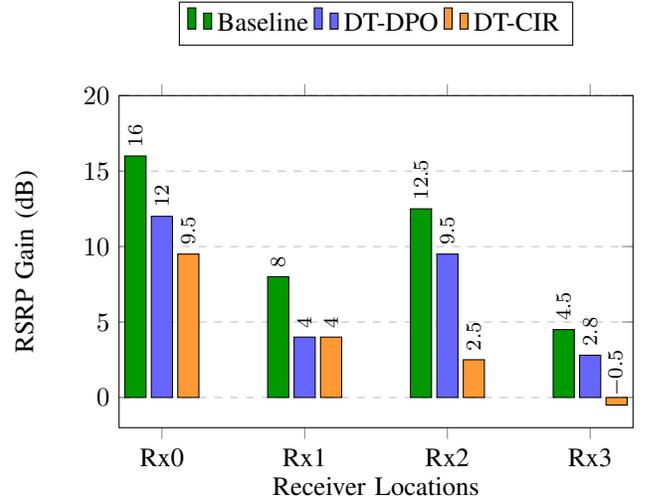

Fig. \ref{fig:all_patterns} compares the phase configuration patterns obtained from the physical iterative search based optimization and the DT for different UE locations. As depicted in the figure, the DT generated patterns show a strong correlation with the physically optimized local optimum patterns.

Fig. \ref{fig:coverage_maps} shows the coverage maps within Sionna to visually demonstrate the potential power gain and observe the impact of the RIS configurations on the propagation environment. Fig. \ref{fig:coverage_maps}(a) illustrates the scenario with an unconfigured RIS (all-zero phases), resulting in a diffuse signal distribution. In contrast, Fig. \ref{fig:coverage_maps}(b) depicts the environment when the RIS is configured using the specific DT-DPO phase pattern illustrated in Fig. \ref{fig:all_patterns}(a). This visualization confirms that the optimized configuration successfully illuminates the intended area (Rx0).

Fig. \ref{fig:rsrp_results} presents the RSRP gains for four different UE locations (Rx0-Rx3). The results indicate that the configurations derived from the proposed DT-DPO and DT-CIR usually provide positive gains, closely following the trend of the physical benchmark. For Rx0, which is likely in a strong reflection path, the physical optimization yields a $16$ dB gain. The proposed DT-based approach achieves a gain of $12$ dB, demonstrating comparable performance to the physical benchmark. 
Also, while the physical optimization requires $N = 128$ operations, our proposed DT-DPO and DT-CIR framework achieves this performance with $|\tilde{\mathcal{X}}_{DT-DPO}| = 2$ and $|\tilde{\mathcal{X}}_{DT-CIR}|=1$ operations, respectively, reducing overhead and latency. Here, an operation is defined as the process of loading a phase configuration onto the RIS controller and measuring the resulting RSRP. 

The slight discrepancies observed between the physical optimization and DT-based optimizations can be attributed to minor geometrical mismatches in the 3D model or hardware impairments not fully captured in the ray tracer. Specifically, the larger deviations observed at locations Rx1 and Rx3 arise because these points fall slightly outside the optimal angular operating range of the RIS prototype. Although the RIS remains operational in these angles, this difference in element patterns at off-boresight angles results in a slight mismatch between the simulated and actual RIS model, leading to inaccurate predictions by the DT. However, despite these factors, the DT-based solutions achieve a RSRP comparable to the physical benchmark.

\section{Conclusion} 
\label{sec:conclusion}
In this work, we proposed a practical DT-driven optimization framework for RIS-assisted wireless communication systems. By using ray tracing and 3D environmental modeling, the proposed approach eliminates high pilot overhead associated with optimization or conventional channel estimation methods in passive RIS deployments. We demonstrated that the optimal RIS phase configurations can be computed entirely within the DT. Our experimental validation confirms that geometry-based DT frameworks offer a feasible and efficient path for the deployment of large-scale RIS in 6G networks. Future work will focus on extending this framework to dynamic scenarios with mobile users, adapting the optimization strategies for MIMO systems, and integrating machine learning to compensate for geometric modeling errors.

\balance
\small
\bibliographystyle{IEEEtran}
\bibliography{main.bib}

\vspace{12pt}
\color{red}
\end{document}